\documentstyle[art12]{article}

\advance\voffset by -1.0in

\def \MSbarbasic {\overline{{\rm MS}}}
\def \MSbar {\ifmmode \MSbarbasic \else $\MSbarbasic$\fi }

\def \GeV { {\ \rm GeV} }

\def \tr {\mathop{\rm tr}}
\def \Ndof {N_{{\rm dof}}}
\def\Atrue{A_{{\rm true}}}

\begin{document}

\rightline{CTEQ NOTE 94/01}

\rightline{hep-ph/9411214}

\bigskip
\centerline{\Large \bf Issues in the Determination of}
\centerline{\Large \bf Parton Distribution Functions}

\bigskip
\begin{center}
   Davison E. Soper
\\
   Institute of Theoretical Science,
   University of Oregon,
\\
   Eugene OR 97403, U.S.A.
\\[0.2in]
   John C. Collins
\\
   Department of Physics,
   Penn State University,
   104 Davey Lab.,
\\
   University Park PA 16802, U.S.A.

\bigskip
ABSTRACT
\smallskip
\end{center}
\begin{quote}
The CTEQ and MRS parton distributions involve a substantial number
($\sim$ 30) of parameters that are fit to a large number ($\sim$
900) of data. Typically, these groups produce fits that represent a
good fit to the data, but there is no substantial attempt to
determine the errors associated with the fits. Determination of
errors would involve consideration of the experimental statistical
and systematic errors and also the errors in the theoretical formulas
that relate the measured cross sections to parton distributions. We
discuss the principles that would be needed in such an error
analysis. These principles are standard. However, certain aspects of
the principles appear counter-intuitive in the case of a large number
of data. Accordingly, we strive to devote careful attention to the
logic behind the methods.
\end{quote}

\bigskip
\leftline{November 1994.
Submitted to hep-ph; not intended for journal publication at
this time.}

\bigskip
\section{Some questions}

Many times, parameters of a theory are determined by minimizing the
$\chi ^{2}$ for the deviation between the theoretical predictions and
experimental data.  An important example is given by global fits
of QCD parton densities.  The number of data is large, about 900,
whereas the number of parameters is much smaller, about 30.
In this note, we explain a number of issues in the interpretation
and use of $\chi ^{2}$.  Although we will do this using, as an example,
the global QCD fits (as done in the CTEQ collaboration), our
considerations are quite general.

This note was prompted by some discussions at
a CTEQ meeting in January 1994.  Among other things,
confusion was caused by the following
somewhat counterintuitive aspect to the use of $\chi ^{2}$ to measure
the goodness of a fit of theory to data, when there are a large
number $\Ndof$ of degrees of freedom (number of data points minus
number of parameters in fit).

Now, it is trivial that the expected value of $\chi ^{2}$ is $\Ndof$,
provided only that the theoretical calculations are correct, and that
the estimates of the experimental errors are valid. Moreover, if the
errors are Gaussian and independent, the standard deviation on the
variations in $\chi ^{2}$ is $\sqrt {2\Ndof}$. (We've heard people
say $\sqrt {\Ndof}$, but our calculations and the Particle Data
Booklet give $\sqrt {2\Ndof}$.) Inexactly Gaussian error
distributions do not change these results qualitatively, but will
change the significance of large deviations of $\chi ^{2}$ from the
expected value of $\Ndof$, and will also change the standard
deviation to a different multiple of
$\sqrt {\Ndof}$.

If one compares two fits to the same data, then a one $\sigma $ change
in the fit parameters corresponds to a change of one unit in
$\chi ^{2}$.  Hence an improvement in a fit that causes a decrease in
$\chi ^{2}$ by a modest number is quite significant, e.g., an
increase of $9$ in $\chi ^{2}$ from its minimum corresponds to a
$3\sigma $ effect.  This occurs even though the standard deviation on
$\chi ^{2}$ is $\sqrt {2\Ndof}$, which is much larger, if $\Ndof$ is
large, e.g., ${\cal O}(10^{3})$. It follows that if $\chi ^{2}=910$
for the CTEQ global fit to 900 data points and $\chi ^{2}=920$ for
the MRS
fit to the same data, then the CTEQ fit
is much superior. (MRS would probably prefer to exchange these two
values of $\chi ^{2}$!) On the other hand each fit, by itself, could
be an excellent fit to the data.  The confusing aspect is that if all
the experiments were repeated and the CTEQ fit to the new data gave
$\chi ^{2}=890$, an improvement of 20 units, then one would {\it not}
say that the fits were significantly better for the new data than for
the old data. (But each of the parameters of a fit to the new data
would normally be within a standard deviation or two of the
parameters of the old fit.)

However, the above statements are only true if the error matrix has been
correctly estimated, and if the formulas used in the theoretical fits
are exactly correct.  Neither of these assumptions is necessarily
true for the global fits, particularly if the errors on different
points are assumed to be independent.  So we will give a discussion
of how to define and use $\chi ^{2}$ in the presence of correlated
experimental errors and of theoretical errors.

Another interesting issue is that there are experiments,
notably direct photon production, that determine rather directly the
gluon density.  But the good precision and the large quantity of the
DIS data appear to render the direct photon data unimportant in the
determination of the gluon density, especially after allowing for a
systematic error on the overall normalization of each experiment.  We
suspect that an improved treatment of systematic errors (experimental
and theoretical) may help here.

Moreover, $\chi ^{2}$ is not the only possible measure of goodness of
fit.  If the theory is correct, and if the errors are estimated
appropriately, then minimizing $\chi ^{2}$ is what one should be doing.
But if one is to decide whether the theory calculations fail to
agree with experiment,\footnote{
   If a theoretical calculation fails to agree with experiment,
   that does not automatically imply that the theory is wrong.
   Even if the fundamental theory (QCD for our example) is
   correct, its implementation in a calculation may be limited in
   precision: for example we have parameterizations for unknown
   functions (the parton density functions), and QCD radiative
   corrections are calculated to a low order in perturbation
   theory.  It goes beyond statistical considerations to decide
   on the physical implications of a disagreement between the
   theoretical calculation and experiment.
}
one should look to additional measures.  One idea is simply to compute
$\chi ^{2}$ for the fit, restricting to some subset of the data.  We
will show how to define the effective number of degrees of freedom
for the subset: the number of data minus the effective number of
theory parameters for the subset.  This number is important for
determining the expected value of $\chi ^{2}$.

A final point was that of how to estimate the error on the QCD
prediction of some quantity, given the errors in the global fit. One
use of such an error estimation is to determine how accurately
$m_{W}$ can be measured at a hadron collider.  Merely trying a small
number of different parameterizations of the parton distribution
functions appears insufficient.  We will present a systematic way to
present the errors on the fits, so that the error on the predictions
can
readily be calculated. One can then imagine going on to determine
the most economical way to improve the data on which the fit is
based, if the predictions are to be improved.


\section{Example}

First, we give a simple example to remind everyone why improving
a fit by one unit in $\chi ^{2}$ is a 1 standard deviation
effect, even though the
standard deviation of $\chi ^{2}$ is about $\sqrt {2N}$.

Consider an experiment that measures a particular cross section and
whose errors are purely statistical.  Let the experiment run for $N$
days, and for simplicity let the integrated luminosity on each day be
the same.  Let $\sigma _{i}$ be the measurement of the cross section
from the running on day $i$, and let it have error $\Delta \sigma $.

Then the measurement of the cross section from the whole $N$ days of
running is
\begin{equation}
  \sigma  = \frac {\sum _{i}\sigma _{i}}{N}.
\label{sigma}
\end{equation}
The error on $\sigma $ is $\Delta \sigma /\sqrt N$, since the errors
on each day are independent.

Now consider measuring the cross section by minimizing
\begin{equation}
  \chi ^{2}(T) \equiv
  \sum _{i=1}^{N} \frac {(\sigma _{i}-T)^{2}}{(\Delta \sigma )^{2}}
\label{chi2T}
\end{equation}
as $T$ is varied.  It can easily be checked that the
minimum of $\chi ^{2}$ is at $T=\sigma $, as defined by Eq.\
(\ref{sigma}):
\begin{equation}
   {(\chi ^{2})}_{{\rm min}} = \chi ^{2}(\sigma ).
\end{equation}
If we consider an ensemble of repetitions of the whole experiment and
fit, then the mean of ${(\chi ^{2})}_{{\rm min}}$ is
$N-1$, and for large $N$ its standard deviation is $\sqrt {2N}$.

In contrast, let us consider just one repetition of the $N$ day
experiment, and ask how much $\chi ^{2}(T)$ changes when one changes $T$
by one standard deviation on the measurement error of the cross
section.  Now
\begin{equation}
\chi ^{2}(T)-{(\chi ^{2})}_{{\rm min}}
  = N \frac {(T-\sigma )^{2}}{(\Delta \sigma)^{2}},
\end{equation}
Thus a change of $T$ from $\sigma $ by $\Delta \sigma /\sqrt N$ gives
$\chi ^{2}$ that is one unit above its minimum value.  That is, the
one-standard-deviation error on the measurement of the cross section
is given by a change of $\chi^{2}$ by one unit from its minimum.


\section{General definition of $\chi ^{2}$}

The theory of $\chi ^{2}$ in the presence of correlated errors is not
always treated in textbooks.  It is treated is in
\cite{Barlow,Eadie}.

\subsection{Experimental errors}

We consider a general situation, with
data from several experiments.  Suppose the total number of
data is $N$.  Label them $E_{i}$ with $i = (1,\dots,N)$.  The true
values are $V_{i}$, but there are experimental errors. We treat
the experimental errors as drawn from a Gaussian distribution:
\begin{equation}
E_{i} = V_{i} + \sum _{J = 1}^{a} M_{iJ}x_{J}\,.
\label{experror}
\end{equation}
Here the $x_{J}$ are independent Gaussian random variables with mean
$\langle x \rangle = 0$ and variance
$\langle x^{2}\rangle -\langle x\rangle ^{2} = 1$.
For statistical errors, the number of random
variables $x$ equals the number of data and the matrix $M$ is
diagonal, with $M_{ii}$ being the expected error on the measurement
of datum $i$. But when {\it systematic} errors are included as well,
there are more
$x$'s, and $M$ is not diagonal.  For measurements of cross sections,
the total number of $x$'s is at least the number of data.
The true values $V_{i}$ are, of course, not necessarily known.

Note that some of the formulae we derive will not depend on the
$x$'s being Gaussian, but only on their independence. For example we
have:
\begin{eqnarray}
   \langle x_{I}x_{J}\rangle  &=& \delta _{IJ} \,,
\\
   \langle E_{i}E_{j}\rangle  &=& V_{i}V_{j} + \sum _{J=1}^{a} M_{iJ}
M_{jJ} \,.
\end{eqnarray}

For statistical errors, the distribution of the errors is known.  If
the $E_{i}$ represent measured cross sections then the statistical
errors are distributed according to the Poisson distribution. As long
as the number of counts represented by each $E_{i}$ is large, the
Poisson distribution may be adequately approximated by a Gaussian
distribution.

For systematic errors, the distribution is not well known.  The
correction for various experimental effects requires the exercise of
judgment on the part of the experimental groups; the error estimate
represents a judgment as to the uncertainty in the corrections.  The
actual distribution of $E_{i} - V_{i}$ due to systematic errors in an
ensemble of high energy physics experiments might conceivably be
measured by looking at past experiments, for which ``correct'' results
$V_{i}$ are now known from more accurate experiments. This thought
makes it clear that the question of the distribution of systematic
errors is as much a question of sociology of science as it is of
science itself.

Despite these difficulties of interpretation, we proceed.  We take
eq.~(\ref{experror}) as a reasonable model for the distribution of
the experimental systematic errors in the regime in which these
errors are small ($\sim 1 \ \hbox{standard deviation}$).
For an estimate of the size of
these errors, the $M_{iJ}$, we have little choice but to take the
experimental groups seriously and thus use their values. Experience
indicates that the probability for the actual error to be many
standard deviations
is small, but is much larger than indicated by a Gaussian
distribution.  Similar remarks apply to errors on the theory, which
we will treat in a moment. Thus after making a fit to parton
distributions, one should  check whether the deviation between any
datum and the theory is greater than 2 or 3 standard deviations.  If
it is, that is a signal that the corresponding experiment and theory
calculation need to be reexamined.

\subsection{Theoretical errors}

For each datum, there is a theoretical prediction $T_{i}(A)$, which
depends on the parameters $A=(A_{1},\dots,A_{P})$  of the theory.  (For
the CTEQ fits, these
are the fundamental parameters of QCD, the parameters of the
parton distributions, etc.)  Suppose the true values of the parameters
are $\Atrue$. Then one might expect that the true values
corresponding to the data equal the results of the theoretical
calculation: $T_{i}(\Atrue) = V_{i}$.

But there are also theoretical errors. For instance, let the
theoretical prediction for the Drell-Yan cross section
$d\sigma /dQ^{2}\, dy$ be $f(Q^{2},y)$. If the calculation is to order
$\alpha _{s}^{2}$, then we might expect that there is a correction due
to the uncalculated higher order corrections that takes the form
\begin{equation}
\delta f_{1} = c_{1}\,  \alpha _{s}^{3}\, x_{1} \times f(Q^{2},y)\,,
\end{equation}
where we treat $x_{1}$ as a Gaussian random variable with variance 1
and, let us say, $c_{1} = 1$. This corresponds to an unknown
``$K$-factor'' that is a constant. In addition, we might expect that
there is another correction due to the uncalculated higher order
corrections that is not constant, but is largest for large rapidities:
\begin{equation}
\delta f_{2} = c_{2}\,  \alpha _{s}^{3}\, y^{2}\, x_{2} \times
f(Q^{2},y)\,,
\end{equation}
where here we might take $c_{2} = 0.1$. Finally, we might suppose that
there is a higher twist correction of the form
\begin{equation}
\delta f_{3} = c_{3}\, {1 \GeV \over Q^{2}}\, x_{3} \times f(Q^{2},y)\,,
\end{equation}
with $c_{3} = 1$. Of course the form and size of the ``theoretical
error'' contributions can and should be debated, although their
existence is indisputable. What is given above is just an example.

Thus we can model theoretical errors in the same way as the
experimental systematic errors, with more Gaussian random variables
$x_{J}$:
\begin{equation}
T_{i}(\Atrue) = V_{i} - \sum _{J = a+1}^{b} M_{iJ}x_{J}\,.
\label{therror}
\end{equation}
We have a minus sign in eq.~(\ref{therror}) instead of the plus
sign eq.~(\ref{experror}) so that the error variables appear in the
formula for $\chi ^{2}$ all with the same sign.

Of course, the use of Gaussian errors for the theoretical errors is
even more problematical than for experimental systematic
errors.\footnote{%
   But one can imagine doing a historical investigation, just as
   one might investigate systematic errors on experiments.
   One could apply current criteria for errors on theory to the
   state of theoretical knowledge some years ago.  Then one could
   ask how valid these error estimates are in the light of more
   accurate later results.
}
Despite the difficulties, we proceed.  We take eq.~(\ref{therror}) as
a reasonable model for the distribution of the theoretical errors in
the regime in which these errors are small. For an estimate of the
size of these errors, the $M_{iJ}$, we use our own judgment combined
with that of the authors of the theoretical papers. Again, we suspect
that the probability for the actual error to be many standard
deviations is
small, but is much larger than indicated by a Gaussian distribution..

\subsection{$\chi ^{2}$ and its interpretation}

We combine eqs.~(\ref{experror}) and (\ref{therror}) to get
\begin{equation}
E_{i} - T_{i}(\Atrue) = \sum _{J = 1}^{b} M_{iJ}x_{J} \,.
\label{EminusT}
\end{equation}
But we do not know $\Atrue$.

Now, given particular values $A$ of the parameters and given
the experimental data $E$, one defines the likelihood ${\cal L}(A,E)$
as the probability (per unit $dE$) that the data $E$ would be
obtained if the parameters' values were $A$. We can calculate ${\cal
L}(A,E)$ by integrating over the random variables $x$. The result is
\begin{equation}
{\cal L}(A,E) \propto  e^{-\frac {1}{2}\chi ^{2}} ,
\label{likelihood}
\end{equation}
where
\begin{equation}
\chi ^{2}(A,E) = \sum _{i,j=1}^{N} \left( E_{i} - T_{i}(A) \right)
               \,{\cal E}^{-1}_{ij}\,
               \left( E_{j} - T_{j}(A) \right) \,,
\label{chi2def}
\end{equation}
with
\begin{equation}
{\cal E}_{ij} = \sum _{K=1}^{b} M_{iK}M_{jK}
\end{equation}
The matrix ${\cal E}$ is the covariance of the deviations of the
data from the true theory:
\begin{equation}
   \Big\langle  \left( E_{i} - T_{i}(\Atrue) \right)
     \left( E_{j} - T_{j}(\Atrue) \right)
   \Big\rangle
   = {\cal E}_{ij} \,.
\label{calE}
\end{equation}
Its diagonal elements give the standard deviation of
the deviation between datum and theory:
\begin{equation}
   \sigma _{i} = \sqrt {{\cal E}_{ii}}.
\end{equation}
Then ${\cal E}^{-1}$ in eq.~(\ref{chi2def}) is a
metric on the space of data.

Notice that Eq.~(\ref{calE}) follows from Eq.~(\ref{EminusT})
and $\langle x_{I}x_{J}\rangle = \delta _{IJ}$.  It is not necessary
that the distribution of the $x$'s be Gaussian.  Although our
formulae for $\cal E$, such as eq.~(\ref{calE}), involve the true
theory parameters, which are not known, $\cal E$ can be and is
estimated from a knowledge of the sources of error alone.

To estimate the correct value of the parameters from the data, one
should choose the parameters $A$ so as to maximize the likelihood
associated with the fit, that is, so as to minimize
$\chi ^{2}$.  The resulting value is a valid estimate provided that
the distribution of the errors is close enough to Gaussian, and that
our estimate of the errors is valid.  The expectation value of
${(\chi ^{2})}_{{\rm min}}$
is the number of degrees of freedom:
\begin{equation}
   \left\langle  {(\chi ^{2})}_{{\rm min}} \right\rangle  = N - P,
\end{equation}
where $P$ is the number of parameters we fit.

Now recall what we said earlier.
For the CTEQ fits, we expect $\chi ^{2}$ to be around
900, since the number of $x$'s minus the number of fit parameters is
about that. The standard deviation in $\chi ^{2}$ is about 40 or 50.
Nevertheless, if it turns out that the best fit has $\chi ^{2} = 914$
then a fit with different parameters with $\chi ^{2} = 918$ is
significantly worse (at the ``$2\sigma $'' level).  This seems
incredible! How can a change in $\chi ^{2}$ of four parts per mill be
significant?

First, the result rests on assumptions about the error distributions,
so the discussion given above about these distributions needs to be
taken seriously.  In particular, if there are correlated errors on
different data points, then these correlations must be taken into
account.  If the correlations are ignored then changes of $\chi ^{2}$
by a few units need not be significant.  For example suppose an
experiment provides a large number $N$ of data points, and that the
overall normalization (from a luminosity measurement) is the main
source of error for each point.  Suppose the true luminosity is
$2\sigma $ away from the assumed value.  This will produce a
contribution of 4 units to $\chi ^{2}$, if the errors are treated
correctly, but a contribution of about $4N$ units, if $\chi ^{2}$ is
calculated on the hypothesis that the errors are uncorrelated.

Suppose we assume that the Gaussian form of distribution of the
$x$'s is valid, and that we have correctly estimated the
correlations. Then we can calculate the probabilities of different
values of the parameters.

Let us examine this question at first based on having just two
possible fits.  Later, we will generalize to having a 25 dimensional
space of fits. We suppose that we have two fits, or models, labeled 1
and 2, and that, somehow, we know that one or the other of them must
be correct. The two fits are, we suppose, similar, but have different
values for the parameters $A$. Furthermore, let us assume that there
is little to choose between models 1 and 2 if we don't look at the
data, so that we judge these fits to be {\it a priori} equally
likely. That is to say, we ascribe probabilities $P_{1}^{(0)} = 0.5$
and $P_{2}^{(0)} = 0.5$ that these distributions are right. (Perhaps
another observer would have a somewhat different judgment.)  Let the
corresponding $\chi ^{2}$'s for the two fits be $\chi ^{2}_{1} = 914$
and $\chi ^{2}_{2} = 918$. After comparing to data, we judge that the
probability that model $n$ is correct ($n=1,2$) to be
\begin{equation}
P_{n}^{\prime }= {P_{n}^{(0)} {\cal L}_{n} \over P_{1}^{(0)}
{\cal L}_{1} + P_{2}^{(0)} {\cal L}_{2}}.
\label{bayes}
\end{equation}
Here the likelihood ${\cal L}_{n}$ is the probability that the
given experimental result is obtained if model $n$ is right, as given
in eq.~(\ref{likelihood}).  These likelihoods are calculated using the
Gaussian distribution for the errors. (The result (\ref{bayes}) is
Bayes' theorem, but it is evidently not very deep from a mathematical
point of view.)  Thus the ratio of the probability that model 1 is
right to the probability that model 2 is right is
\begin{equation}
{ P_{1}' \over P_{2}'} =
   e^{- {1\over 2}(\chi ^{2}_{1} - \chi  ^{2}_{2})}\times
   { P_{1}^{(0)} \over P_{2}^{(0)}}
=
   e^{2}\times {P_{1}^{(0)} \over P_{2}^{(0)}}\sim 10\,{ P_{1}^{(0)} \over
   P_{2}^{(0)}}.
\end{equation}
Recall that we assumed that ${ P_{1}^{(0)} / P_{2}^{(0)}}\sim 1$, so that
${ P_{1}'/ P_{2}'} \sim 10$.

In some instances, one would judge the
two models to have a different initial probability. For instance if
model 2 corresponds to the LEP value of
$\Lambda _{\MSbar}$ while model 1 corresponds to a value that
differed by $2\sigma _{{\rm LEP}}$, then model 2 would be initially
favored by a factor of about 10 and after examining the global fit one
would judge models 1 and 2 to be equally likely. One would also be a
bit concerned about why there was a $2\sigma $ disagreement between
electron-positron experiments and experiments involving hadrons.

Notice, for example, that a $\chi ^{2}$ difference of 40 corresponds
to a likelihood ratio of $\exp(20)\sim 10^{10}$, an overwhelming
difference, even though 40 doesn't seem like much in comparison with
$\sqrt {2N}$, the standard deviation of ${\chi ^{2}}_{{\rm min}}$.
Nevertheless, if one produced a fit with a $\chi ^{2}$ of 940 for
900 degrees of freedom, one would not consider that grounds for
regarding  the fit as bad, even though $\chi ^{2}$ is 40 above its
expectation value.

In Sect.~\ref{sec:ErrorsInParams}, we will generalize to having a
space of fit parameters instead of just two possible models toward
the end of the next section, after discussing $\chi ^{2}$ as a
function of the fit parameters.  Let us just note here that the true
value of any one parameter in the fit is not likely to be more than
$2\sigma $ from the value determined by the best fit.  However, there
are many fit parameters, roughly 25 of them.  Each of them is likely
to be $1\sigma $ away from the fitted value.  For this reason, $\chi
^{2}$ for the true parton distribution is likely to be of roughly 25
greater than the $\chi ^{2}$ for the best fit.

\subsection{Non-Gaussian errors}

The precise estimates of probabilities and likelihoods depend on the
assumption that the errors are Gaussian.  Of course, small changes
from a Gaussian distribution will not matter much.  However, large
deviations are another matter.

For example, suppose we have such strongly non-Gaussian errors that the
4th moment of the distribution of one of the errors does not
exist. For example, the probability density of one of the variables
$x$ might go like $1/x^{3}$.  This does not seem to be totally
impossible, at least if one regards a large number for the 4th moment
as being infinity for practical purposes.  In such a situation, the
standard deviation of $\chi ^{2}$ is infinite.  Could this actually be
happening for some systematic and theoretical errors?

\subsection{How much computation is needed?}

An important issue is whether it takes too much computational
effort to use the correct definition of $\chi ^{2}$ with correlated
errors, since rather large matrices are involved.

Let $b$ be the number of error variables $x_{i}$.  Then
the construction of $\chi ^{2}$ involves the
multiplication of an $N\times b$ matrix by a $b\times N$ matrix, to
obtain ${\cal E}$, followed by the inversion of the $N\times N$ real
symmetric matrix ${\cal E}_{ij}$.
This requires at most
about $N^{2}(2b+cN)$ floating point operations, where $c$ is a
constant of order 1.

For the CTEQ fits, $N\approx 1000$, and $b$ is somewhat larger, but
probably by less than a factor of two.  Optimizations using the
symmetry of ${\cal E}$ and the diagonality of the part of $M_{iJ}$
that refers to statistical errors can reduce the number of operations
somewhat.  The calculation involves storing the $N^{2}$ matrix
elements and performing about $3N^{3}$ floating point operations.
This appears to be within the capabilities of UNIX workstations, such
as are used by CTEQ.  Although the calculation of
${\cal E}^{-1}$ might require tens of minutes, it only needs to be
done only once for a whole series of fits.  Recall that producing a
set of CTEQ fits by minimizing $\chi ^{2}$ requires tens of hours.  If
one omits the optimization of using the same ${\cal E}^{-1}$ for a
whole series of values of the parameters, then the computational
load can easily become prohibitive.

A more significant computational load is from the calculation of
$\chi ^{2}$ from its definition (\ref{chi2def}).  This requires
$N^{2}$ floating-point operations instead of the $3N$ needed with
uncorrelated errors, but this calculation is repeated every time a
new set of parameters is used for the theory.  However the
calculation of each of the $N$ theory values $T_{i}(A)$ involves a
lot of calculation, particularly if higher order QCD corrections are
used.  Moreover the evolution of the parton densities for each new
set of parameters is computationally expensive: $10^{6}$ or more
floating point operations.  Thus, the number of operations to
calculate $\chi ^{2}(A)$ for one set of parameters is
$2N^{2}+T_{{\rm thy}} N+T_{{\rm evolve}}$, where $T_{{\rm thy}}$ is
the number of operations to calculate one theory point, and
$T_{{\rm evolve}}$ is the number of operations to evolve a set of
parton densities from an initial value.  Although the $N^{2}$ term
will dominate for asymptotically large $N$, it probably does not
dominate for $N\approx 1000$ in the present CTEQ global analysis. If
we needed, we could also apply some thought to the problem to reduce
the effective value of $N$.

Another issue in the computation is the amount of memory used.
When the errors are uncorrelated, the memory used in the
calculation of $\chi ^{2}$ is proportional to $N$.  But when the errors
are correlated, we need to store one or two $N\times N$ matrices.  This
implies many megabytes of storage, and somewhat less if we rely
on the symmetry of the matrices.  This is within the capabilities
of UNIX workstations, but only if they are suitably equipped.

\section{The errors in the fit parameters}
\label{sec:ErrorsInParams}

In this section we review the general result for the
one-standard-deviation errors on the fit parameters, and show how
these can be used to obtain the errors on further theoretical
predictions.

\subsection{Error matrix on parameters}

The CTEQ fits to parton distribution are functions of a number $P$ of
parameters $A_{\alpha }$. Let $A_{\alpha }^{(0)}$ be the parameters
corresponding to the best fit and let $\delta \! A_{\alpha } \equiv
A_{\alpha } - A_{\alpha }^{(0)} $. Expanding $\chi ^{2}$ to second
order in the $\delta \! A_{\alpha }$, we have
\begin{equation}
   \chi ^{2} \approx  \chi ^{2}_{\rm min}
        +\sum _{\alpha ,\beta =1}^{P} E_{\alpha \beta }^{-1}\,
                 \delta \! A_{\alpha }\, \delta \! A_{\beta } \,,
\label{chi2quad}
\end{equation}
where $E^{-1}_{\alpha \beta }$ is a real, symmetric $P\times P$
matrix. Its inverse, $E_{\alpha \beta }$, is denoted the error matrix
for the parameters.
\footnote{
   Note the distinction between $E_{\alpha \beta }$, which concerns the
errors
   on the fit parameters, and ${\cal E}_{ij}$, which concerns the
   errors on the data points.  Note also that we have
   ``overloaded'' the symbol $E$: $E_{\alpha \beta }$ is the
   error matrix, while $E_{i}$ denotes data values.
}
This equation provides a practical way to calculate the error matrix
since, in the process of minimizing the $\chi ^{2}$ of a fit to data,
one obtains enough information to calculate the second derivatives of
$\chi ^{2}$ with respect to the theory parameters $A_{\alpha }$.

We can gain more insight into $E_{\alpha\beta}$ if we relate it to
the error matrix ${\cal E}_{ij}$ associated the data. For this
purpose, we consider the dependence of the theoretical predictions
$T_i(A)$ on the parameters $A$. Here, we make an approximation. We
replace the $T_i(A)$ by linear functions,
\begin{equation}
T_i(A) \approx T_i(\Atrue) + \sum_j
(A_j - A^{\rm true}_j)\nabla_j T_i(\Atrue ).
\label{linearT}
\end{equation}
There is a simple intuitive justification for this. The important
region in which we need an accurate representation of the $T_i(A)$ is
that in which $T_i(A)$ is within about one experimental standard
deviation of $T_i(\Atrue)$. This is a small region, $\Delta T/T\sim $a
few percent. In this region, the linear approximation (\ref{linearT})
is adequate as long as we have chosen a decently smooth
parameterization of the parton distributions. In the arguments that
follow, one could extend eq.~(\ref{linearT}) to a quadratic
dependence, producing new terms proportional to $\nabla_i\nabla_j T$
and factors like $T_i-E_i$. We have not done this because there is a
cost in complicating the formulas.

We apply the approximation (\ref{linearT}) to compute the error
matrix $E_{\alpha\beta}$. Expanding now about the best fit values
$A^{(0)}$ of the parameters, we write
\begin{equation}
T_i(A) \approx T_i(A^{(0)}) + \sum_\alpha
\delta A_\alpha \nabla_\alpha T_i(A^{(0)} ),
\label{linearTencore}
\end{equation}
where, as in Eq.~(\ref{chi2quad}),
\begin{equation}
\delta A_\alpha = A_\alpha - A^{(0)}_\alpha.
\end{equation}
Inserting this into Eq.~(\ref{chi2def}) for $\chi^2$, we obtain
\begin{eqnarray}
   \chi ^{2}(A) &\approx & \chi ^{2}_{\rm min} +
           \sum _{ij\alpha \beta } {\cal E}^{-1}_{ij} \,
                 \nabla _{\alpha }T_{i} \, \nabla _{\beta }T_{j} \,
                 \delta \! A_{\alpha } \, \delta \! A_{\beta }
\nonumber \\
          &=& \chi ^{2}_{\rm min} +
             {\delta \!A}^{T} \, \nabla T^{T} \,
             {\cal E}^{-1} \nabla T \, \delta \! A \,,
\label{chi2deriv}
\end{eqnarray}
where in the second line, we have used a matrix
notation, with a superscript ${}^{T}$ denoting a matrix transpose.
Eq.~(\ref{chi2deriv}) gives us a formula for the error matrix:
\begin{equation}
   {E^{-1}}_{\alpha \beta } =  \sum _{ij} \,
        {{\cal E}^{-1}}_{ij} \,
        \nabla _{\alpha }T_{i} \, \nabla _{\beta }T_{j} .
\label{ErrorMatrix}
\end{equation}

\subsection{Interpretation of $E_{\alpha\beta}$}

Using the linearity assumption (\ref{linearT}), we find that
the expectation of the deviation of the fit parameters $A^{(0)}$
from their true values $A^{{\rm true}}$ is zero:
\begin{equation}
   \left\langle
     \left( A_{\alpha }^{(0)} - A_{\alpha }^{{\rm true}} \right)
   \right\rangle
   = 0,
\label{MeanDev}
\end{equation}
Furthermore, the covariance matrix of these deviations is equal to
precisely the error matrix:
\begin{equation}
   \left\langle
     \left( A_{\alpha }^{(0)} - A_{\alpha }^{{\rm true}} \right)
     \, \left( A_{\alpha }^{(0)} - A_{\alpha }^{{\rm true}} \right)
   \right\rangle
   = E_{\alpha \beta } .
\label{CovDev}
\end{equation}

One standard way to interpret $E_{\alpha\beta}$ is simply to state
that it is the matrix that appears in Eq.~(\ref{CovDev}). An
alternative interpretation makes use of the Bayesian approach used
earlier in this paper. We recall that
\begin{equation}
{\cal L}(\delta A) \propto \exp(- {\scriptstyle { 1 \over
2}}\chi^2(A))
\propto \exp(-  {\scriptstyle { 1 \over 2}}
\delta A^T E^{-1}\delta A)
\end{equation}
is the probability that the measured experimental results would be
obtained if the parameters had the values $A^{(0)} + \delta A$.
Alternatively, given our knowledge of the experimental results, we
judge the probability that the parameters have the values $A^{(0)} +
\delta A$ to be
\begin{equation}
dP = {\cal N} \exp(-  {\scriptstyle { 1 \over 2}}
\delta A^T E^{-1}\delta A)\, P^{(0)}(\delta A)
\prod_\alpha (d \delta A_\alpha)
\end{equation}
where ${\cal N}$ is the normalizing factor such that $\int dP = 1$
and $P^{(0)}(\delta A)$ is the {\it a priori} probability that
states our judgment of the probability before we know of the
experimental results. We will discuss in Section \ref{sec:Inclusion}
what happens if one includes information based on prior experiments
in $P^{(0)}(\delta A)$. Here we will assume that we lack any
substantial {\it a priori} information, so that $P^{(0)}(\delta A)
\approx const.$ for reasonably small values of $\delta A$. With this
understanding, we can interpret
\begin{equation}
dP = {\cal N}^\prime \exp(-  {\scriptstyle { 1 \over 2}}
\delta A^T E^{-1}\delta A)\,
\prod_\alpha (d \delta A_\alpha)
\label{Aprobability}
\end{equation}
as the probability that the parameters have the values $A^{(0)}
+\delta A$.


\subsection{Errors on predictions}

Consider a physical quantity $S$ the calculated value of which
depends on the parton distributions (and on any of the other
parameters that we fit). For instance $S$ might be a calculated cross
section. It is a function $S(A)$ of the parton parameters. The best
value for this physical quantity is $S(A^{(0)})$.  But what error
arising from the parton distributions should be ascribed to this
result? And how can one publish parton distributions that would make
it easy to calculate this error?

Using Eq.~(\ref{Aprobability}),the probability that the quantity in
question takes the value $S + \delta S$ is
\begin{eqnarray}
{\cal L}(\delta S) &\propto &
\int \prod _{\alpha } \left( d\,\delta \! A_{\alpha }\right)\
\exp(-  {\scriptstyle { 1 \over 2}}
\delta A^T E^{-1}\delta A)\
\delta \!\left(
\delta S - \sum  \nabla _{\alpha }S \delta \! A_{\alpha }
 \right).\nonumber\\ &\propto &
\exp\left( - { (\delta S)^{2} \over 2 \sum \nabla _{\alpha }S
E_{\alpha \beta } \nabla _{\beta }S}
\right).
\end{eqnarray}
Thus we obtain a Gaussian probability distribution for $S$ with an
standard error
\begin{equation}
\sigma _{S} =
\sqrt {\sum  \nabla _{\alpha }S\ E_{\alpha \beta }\, \nabla _{\beta
}S}.
\label{Serror}
\end{equation}

\subsection{Parton distributions with errors}

This result suggests a fairly simple strategy.  One can publish the
matrix $E_{\alpha \beta }$ together with $P+1$ parton distributions,
one corresponding to the best fit and each of the others corresponding
to a small change $\delta \!A_{\alpha }$ in {\em one} of the
parameters $A_{\alpha }$. This enables the user to calculate $\nabla
_{\alpha }S = \partial S(A)/\partial A_{\alpha }$. Given the $\nabla
_{\alpha }S$ and the matrix $E_{\alpha \beta }$, the user can
calculate
$\sigma _{S}$, from eq.~(\ref{Serror}).  This result is important: it
correctly represents the uncertainty in the prediction of $S$.

An alternative might be to diagonalize the error matrix, and to give
a set of eigenvectors $B^{(a)}_{\alpha }$, with $a=1, \dots, P$.  The
normalization of each eigenvector would be such that it corresponds
to a $1 \, \sigma $ change in the parameters:
\begin{equation}
   \chi ^{2}(A^{(0)}+ B^{(a)}) = \chi ^{2}_{\rm min} + 1
\end{equation}

It may seem like too much to distribute something like 26 parton
distribution functions.  But really 26 is not an enormous number,
at least if one publishes the distributions electronically in the
form of interpolating tables, for example. In
fact, the present CERN package of all parton distribution functions
contains something like this number of parton distribution sets.  It
seems clear that a set of 26 parton distributions that would enable
users to calculate an honest error in any quantity of interest is
vastly superior to a set of 26 old and new parton distributions that
are not related to one another in any coherent way.  Furthermore, all
information on errors is contained in these distributions.


\section{Inclusion of expectation on $A$}
\label{sec:Inclusion}

When performing a fit to data, we may already have {\it a priori}
knowledge about likely values of the parameters.  This knowledge
might be from a fit to data from other experiments not included in
the CTEQ fits, for example a measurement of the scale $\Lambda $ from
LEP data. Another situation is that we might have
parton distributions based on a fit to
an ``old'' set of data, but might not
have access to the old data and details of its errors. Another
possibility is that we might have new data from one experiment, and
wish to avoid the computational load of doing a fit to the combined
set of old and new data.   In any of these situations, we can make
use of our knowledge of the errors on the parameters $A$ if the
errors in the old fit are uncorrelated with the errors in our new
fit. (Or, at least, if this is a reasonable approximation.)

Since likelihoods multiply, one can include the information on the
parameters $A$ in a fit by adding a term to our definition
eq.~(\ref{chi2def}):
\begin{eqnarray}
\chi ^{2}(A,E) &=& \sum _{i,j=1}^{N} \left( E_{i} - T_{i}(A) \right)
               \,{\cal E}^{-1}_{ij}\,
               \left( E_{j} - T_{j}(A) \right)
\nonumber\\
&&+
\sum _{\alpha ,\beta =1}^{P}
\left( A_{\alpha } - A^{\rm old}_{\alpha } \right)
\,(E^{-1}_{\rm old})_{\alpha ,\beta }\,
\left( A_{\beta } - A^{\rm old}_{\beta } \right).
\label{chi2newdef}
\end{eqnarray}
Using this new definition is equivalent to fitting the whole set of
new and old data, provided that the errors in the old fit, represented
by $E_{\rm old}$, are uncorrelated with the new errors represented by
${\cal E}$.  One should check that new fit parameters do not deviate
by a large amount from the old parameters, If there are large
deviations, we have an inconsistency which must be investigated.

The distribution of the deviations of the fit parameters from
their true values gets modified, as does the average value of
$\chi ^{2}$.  The mean values of the fit parameters continue to be the
true values, eq.~(\ref{MeanDev}), but the covariance matrix is
changed to
\begin{equation}
   \left\langle
     \left( A_{\alpha }^{(0)} - A_{\alpha }^{{\rm true}} \right)
     \, \left( A_{\alpha }^{(0)} - A_{\alpha }^{{\rm true}} \right)
   \right\rangle
   = E_{\alpha \beta }^{{\rm new}} ,
\label{CovDevNew}
\end{equation}
where
\begin{equation}
   E_{{\rm new}}^{-1} = E^{-1} + E_{{\rm old}}^{-1},
\label{ENew}
\end{equation}
and $E_{\alpha \beta }$ is defined by the same formula as before,
eq.~(\ref{ErrorMatrix}).

Notice that the addition of the extra term to $\chi ^{2}$ changes its
expectation value from $N-P$ to $N$, which is made up of a
contribution
\begin{equation}
N -\tr
\left[
(1 + E_{\rm old}^{-1}\,E )^{-1}
\right] ,
\end{equation}
from the old definition, eq.~(\ref{chi2def}), of $\chi ^{2}$, and a
contribution
\begin{equation}
   \tr \left[
          (1 + E_{\rm old}^{-1}\,E )^{-1}
    \right] ,
\end{equation}
from the extra term that gives the information on the old fit.

\subsection{Determining experimental normalizations, etc}

Another application of the same idea is the measurement of systematic
error parameters for the experiments.  An obvious example is a poorly
known integrated luminosity for a particular experiment whose
statistical precision is good.  In our formalism so far, the error on
the luminosity corresponds to one of the random error variables
$x_{K}$ in eq.~(\ref{experror}).  We could remove this error from the
list of experimental errors, and instead add a theory parameter $f$,
which would be a scaling factor for the theoretical prediction of
data for the experiment. Thus, in eq.~(\ref{chi2def}), we would
replace $E_{i}-T_{i}$ by $E_{i}-fT_{i}$ for each of the data from the
experiment.  We would also add a term
\begin{equation}
   \frac {(f-1)^{2}}{\sigma _{f}^{2}}
\end{equation}
to $\chi ^{2}$, with $\sigma _{f}$ representing the
experiment's estimate of the fractional error on the integrated
luminosity.\footnote{
   Compare \cite{Agostini}, where it was noted that certain
   treatments of normalization errors can systematically bias
   fits in one direction from the data.
}

If a fit to the rest of the data gave an accurate estimate of the
normalization of the cross section, then we would gain a measurement
of $f$; effectively our fit has given a luminosity monitor for the
experiment.  Relative sizes of cross sections in the experiment could
still make an important contribution to the fit, and the improved
knowledge of the luminosity would contribute to other measurements
made by the same experiment.

If, on the contrary, the rest of the data used in the fit are unable
to provide a normalization, then the fit would simply give $f=1\pm
\sigma _{f}$, which simply reproduces our previous knowledge.

One can easily conceive of many variations on this theme.

\subsection{Sum rules}

One especially interesting variation on the theme of fitting
parameters and imposing {\it a priori} knowledge concerns the sum
rules.  Consider, as an example, the function $M$ that defines the
momentum sum rule,
\begin{equation}
   M = \sum _{i} \int _{0}^{1} dx \, x f_{i}(x).
\end{equation}
Here, the sum is over all flavors of quark, antiquark, and
gluon.

The sum rule $M = 1$ follows exactly from the definition of the
parton distributions.  When we want to give the best parton
distributions that we can, we therefore impose $M = 1$ as a
constraint on the parameters $A$. However, it is also of interest
to use the global fit to {\it test} the sum rule, and thus test QCD.
To do this, we can simply let $M$ be one of the parameters $A$ and
fit it as part of a global fit. This would produce a value of
$M$ with errors, which would be of considerable interest.

If ``$M=1$'' or one of the other sum rules does not turn out to be
valid within the errors, we would have something to think about.
Assuming that the sum rules are verified, we could then impose them
as exact constraints and fit again in order to get the ``best''
parton distributions.


\section{$\chi ^{2}$ for subsets of data}

If one admits that the theory may be wrong, or at least that the
approximate calculations based on it are wrong, then the overall
$\chi ^{2}$ is not the only useful measure of goodness of fit.  To
compare the likelihoods of two different theories (or calculations)
it is sufficient to compare their overall $\chi ^{2}$.  But if we want
to know whether a single calculation provides a good fit to the data
we must look further.  Suppose for a fit of certain data we get a
minimum $\chi ^{2}$ of 940 for 900 degrees of freedom.  There is
excess of 40 over the expected value $\langle {\chi ^{2}}_{{\rm
min}}\rangle$. By itself, this value of ${\chi ^{2}}_{{\rm min}}$
represents a good fit, since the standard deviation of $\chi ^{2}$ is
$\sqrt {1800}$, and if the excess of 40 were distributed over many of
the data, then the fit would indeed be good. But if the excess all
came from one point, a 6 standard deviation effect, we would have a
very improbable situation: the fit would be poor, and further
investigation would be warranted.  One case might be an unforeseen
narrow resonance. The statistical issues of resonance searches and
bump hunting are of course quite well-known.

One possibility is to compute $\chi ^{2}$ for a subset of the data
\begin{equation}
   \chi ^{2}(A,E,S) = \sum _{i,j\epsilon S}
               \left( E_{i} - T_{i}(A) \right)
               \,{\cal E}^{-1}_{ij}\,
               \left( E_{j} - T_{j}(A) \right) \,,
\label{chi2defsub}
\end{equation}
where $S$ now labels some subset of the data.  For
example, one might choose $S$ to be the set of all Drell-Yan data, or
the set of all direct-photon data.  The question we now address is how good
the fit to the subset $S$ is.
If the fit is not within expectations for each subset of the data,
then there is something to think about.

Let $N_{S}$ be the number of data in $S$.  If the theory parameters
were fixed at their true values $\Atrue$, and if the errors in the
subset were uncorrelated, then the expectation value of $\chi
^{2}(\Atrue,E,S)$ would be $N_{S}$.  But if $A$ is set to the overall
best fit values $A^{(0)}$, then the expectation value will be less.
An extreme case would be of a CTEQ fit with, say, $N=900$ data and
$P=25$ parameters, and where a subset of 25 data which provided
essentially all of the information in the fit. Then the $\chi ^{2}$
for the subset would be zero, and the remainder of the full $\chi
^{2}$ would be a test of QCD: it should be around 875.

We can use the formalism we have explained to compute the expectation
value of $\chi ^{2}$ for the subset, and thereby determine a measure
of goodness of fit.  In a matrix notation, eq.~(\ref{chi2defsub}) is
\begin{equation}
   \chi ^{2}(A,E,S) = \left( E- T(A) \right)^{T}
               \, P_{S} \, {\cal E}^{-1} \, P_{S} \,
               \left( E - T(A) \right) \,,
\end{equation}
where $P_{S}$ represents a projector onto the subset
of data.  Then the expectation value is
\begin{equation}
   \langle \chi ^{2}(A^{(0)},E,S)\rangle  =
   \tr \left( P_{S} \, {\cal E}^{-1} \, P_{S} \, {\cal E} \right)
   - \tr \left( P_{S} \, {\cal E}^{-1} \, P_{S}
                \, \nabla T \, E (\nabla T)^{T}
          \right) .
\end{equation}
The first term is the effective number of data, which will be $N_{S}$
if the errors in the subset $S$ are uncorrelated with the errors
elsewhere, so that the commutator $[P_{S},{\cal E}]=0$. The second
term represents the effective number of parameters fit by the subset
of data.

If the errors are uncorrelated between different subsets of the data,
$[P_{S},{\cal E}]=0$, then  this interpretation works nicely.  When one
sums the effective number of parameters fit by each subset $S$ over a
complete set of (disjoint) subsets, the total will equal the number
$P$ of parameters.

\section*{Acknowledgements}

We would like to thank colleagues for discussions, in particular,
Steve Heppelmann.

\end{document}